\documentclass[varenna,cite]{cimento}

\input psfig.sty

\title{Condensed matter physics with trapped atomic Fermi gases}

\author{H.T.C. Stoof \atque M. Houbiers}

\institute{Institute for Theoretical Physics, University of Utrecht, 
           Princetonplein 5, 3584 CC Utrecht, The Netherlands}

\PACSes{\PACSit{03.75.Fi}{Phase coherent atomic ensemble 
                             (Bose-Einstein condensation)}
        \PACSit{74.20.Fg}{BCS theory and its development}}

\begin{document}

\maketitle

\begin{abstract}
We present an overview of the various phase transitions that we anticipate to 
occur in trapped fermionic alkali gases. We also discuss the prospects of 
observing these transitions in (doubly) spin-polarized $^6$Li and $^{40}$K 
gases, which are now actively being studied by various experimental groups 
around the world. 
\end{abstract}

\section{Introduction}
After the formidable achievement of Bose-Einstein condensation in spin-polarized 
alkali gases \cite{JILA1,Rice1,MIT1,JILA2,Rice2,MIT2}, the next challenge that 
experimentalists have already set themselves is to realize quantum degenerate 
conditions also in fermionic alkali vapors. One particular motivation in this 
respect is the prediction that a gas of spin-polarized atomic $^6$Li becomes 
superfluid at densities and temperatures comparable with those at which the 
Bose-Einstein experiments are performed \cite{henk1}. As a result of this 
experimental interest, the first theoretical studies of an ideal Fermi gas 
trapped in a harmonic external potential have recently appeared 
\cite{rokhsar,hartmund}. Moreover, the effects of an interatomic interaction 
have also been considered \cite{oliva,marianne1,keith}. It is interesting to 
note that Oliva's calculations for atomic deuterium were already performed a 
decade ago, even though magnetically trapped deuterium had not been observed at 
that time. In fact, it has still not been observed, because the loading of the 
trap cannot be accomplished in the same way as for atomic hydrogen \cite{H1,H2}. 
This is presumably caused by the fact that deuterium binds more strongly to a 
superfluid helium film, that the surface recombination rate is much larger, and 
that the sample is contaminated with atomic hydrogen \cite{meritt}. Fortunately, 
such problems do not arise for experiments with alkali gases and both $^6$Li and 
$^{40}$K have indeed been trapped already \cite{randy,massimo}.

The most important qualitative feature of a trapped Fermi gas is that its 
density profile `freezes' at low temperatures. This is a result of the Pauli 
exclusion principle and can be easily understood by considering an ideal gas in 
the trapping potential $V^{\rm trap}({\bf x}) = m\omega^2{\bf x}^2/2$. In 
general the extent of the gas cloud is determined by the classical turning point 
of the most energetic particles. If the gas is fully classical, i.e., it obeys 
Maxwell-Boltzmann statistics, these particles have an energy of a few $k_BT$ and 
the size of the cloud $R_{TF}$ follows from $m\omega^2 R_{TF}^2/2 \simeq k_BT$, 
implying that $R_{TF} \simeq (2k_BT/m\omega^2)^{1/2}$. We see that, as we lower 
the temperature, the size of the cloud shrinks. Moreover, if we keep the number 
of particles $N$ fixed, the density increases. This process gradually continues  
until we reach zero temperature and the density profile becomes equal to
$n({\bf x}) = N \delta({\bf x})$. If the gas obeys Fermi-Dirac statistics, 
however, the most energetic particles have in the degenerate (nonclassical) 
regime an energy that is equal to the Fermi energy $\epsilon_F$ and the size of 
the gas cloud is always given by $R_{TF} \simeq (2\epsilon_F/m\omega^2)^{1/2}$ 
for temperatures $T \ll \epsilon_F/k_B$. Comparing this also with the density 
profile for an ideal Bose gas, which in the degenerate regime consists of a 
large and narrow condensate peak with a width of about $(\hbar/m\omega)^{1/2}$ 
on top of a broad thermal background of size $R_{TF} \simeq 
(2k_BT/m\omega^2)^{1/2}$, we conclude that the density profile of a degenerate 
Fermi gas is indeed `frozen'. In contrast to the ideal Bose gas, the ideal Fermi 
gas also does not have a phase transition. From a point of view of condensed 
matter physics, an atomic Fermi gas thus appears much less interesting than a 
Bose gas. The main objective of this contribution is to argue that this is no 
longer true if there are interactions between the atoms. 

\section{Interactions}
For atomic alkali gases the most important interatomic interaction is the 
so-called central interaction $V^c(r)$, which consists of a sum of the usual 
singlet and triplet interactions. More precisely, we have that
\begin{equation}
V^c(r) = V_S(r) {\cal P}^{(S)} + V_T(r) {\cal P}^{(T)}~,
\end{equation}
where ${\cal P}^{(S)}$ and ${\cal P}^{(T)}$ denote the projection operators on 
the subspace of singlet and triplet states, respectively. Besides this 
interaction that is the net result of the Coulomb attractions and repulsions 
between the electrons and nuclei of the atoms, we also have to consider the weak 
magnetic dipole-dipole interactions. Of these, the electron-electron magnetic 
dipole interaction is most important and obeys \cite{henk2}
\begin{equation}
V^d({\bf r}) 
 = \frac{\mu_0\mu_e^2}{4\pi} 
   \frac{\mbox{\boldmath $\sigma$}_1 \cdot \mbox{\boldmath $\sigma$}_2 
         - 3 (\mbox{\boldmath $\sigma$}_1 \cdot {\bf \hat{r}})
             (\mbox{\boldmath $\sigma$}_2 \cdot {\bf \hat{r}})}{r^3}
 \equiv - \frac{\mu_0\mu_e^2}{4\pi r^3} \sqrt{\frac{4\pi}{5}}
     \sum_{\mu} (-1)^{\mu} Y_{2\mu}({\bf \hat{r}}) \Sigma^{ee}_{2,-\mu}~.
\end{equation}
Here $\mu_e$ is the electron magnetic moment and the tensor operator 
$\Sigma^{ee}_{2,-\mu}$ is obtained from coupling the Pauli spin matrices 
{\boldmath $\sigma$}$_1$ and {\boldmath $\sigma$}$_2$ of the two valence 
electrons to a tensor of rank 2. 

It should be noted that both these interactions do not commute with the electron 
spin operators $\hbar${\boldmath $\sigma$}$_i/2$ and therefore also do not commute with the atomic hamiltonian, which in a magnetic field contains both a hyperfine and a Zeeman term. As a consequence the central and dipole-dipole interactions are not fully diagonal in the basis in which the atoms are in definite hyperfine states. This is important in principle, because it implies that two atoms can also collide inelastically, i.e., their hyperfine states can change during the collision. Together with three-body recombination events, these inelastic collisions in fact always seriously limit the lifetime of a trapped alkali gas. Nevertheless, we will in the following mostly neglect the nondiagonal parts of the interatomic interaction by restricting ourselves to (doubly) spin-polarized gases for which the `good' elastic collisions dominate the `bad' inelastic ones. Clearly such a restriction is a minimum requirement for our discussion to be also of some experimental interest.

\section{Cooper pairing}
As we will see in detail below, the most common phase transition in a 
weakly-interacting Fermi gas is due to the formation of so-called Cooper pairs. 
The main idea behind the famous Bardeen-Cooper-Schrieffer theory for this 
phenomenon is in fact a Bose-Einstein condensation of these pairs \cite{BCS}. To 
see how we can arrive at a mean-field theory for this phase transition, let us  first recapitulate the mean-field (Hartree) theory for Bose-Einstein 
condensation in an atomic Bose gas. At zero temperature we can then use 
a variational many-body wave function in which all the atoms are in the same 
one-particle state $\phi({\bf x})$, so
\begin{equation}
\Psi_N({\bf x}_1, \cdots ,{\bf x}_N) = \prod_{i=1}^N \phi({\bf x}_i)~.
\end{equation}
In the language of second quantization this essentially reads 
\begin{equation}
|N\rangle = \frac{ \left( \displaystyle \int d{\bf x}~ 
                          \phi({\bf x}) \psi^{\dagger}({\bf x})
                                               \right)^N}{N!}~|0\rangle~,
\end{equation}
where $\psi^{\dagger}({\bf x})$ is the creation operator for an atom at postion 
${\bf x}$ and $|0\rangle$ is the `vacuum' state in which there are no atoms 
present in the trap. Calculating the average of the hamiltonian $H$ in this 
variational wave function and minimizing with respect to $\phi({\bf x})$ leads 
of course to the Gross-Pitaevskii equation \cite{lev} if $N \gg 1$. 

The above variational wave function has a definite number of particles. For 
practical calculations, in particular if we want to consider also nonzero 
temperatures or corrections to the mean-field theory, it is much more 
convenient to consider a variational wave function in which not the number of 
particles $N$, but instead the phase $\vartheta$ is fixed. This state is given 
by
\begin{equation}
|\vartheta\rangle =
  \sum_N \frac{ \left( \displaystyle \int d{\bf x}~ \phi({\bf x})e^{i\vartheta}
                       \psi^{\dagger}({\bf x}) \right)^N}{N!}~|0\rangle
  = \exp \left( \int d{\bf x}~ \phi({\bf x})e^{i\vartheta}
                      \psi^{\dagger}({\bf x}) \right) |0\rangle~,
\end{equation}      
as can be seen from the fact that now the matrix element
$\langle\vartheta|\psi({\bf x})|\vartheta\rangle = \phi({\bf x})e^{i\vartheta}$
has a definite phase in contrast to the matrix element
$\langle N|\psi({\bf x})|N\rangle = 0$. 

It is not difficult to show that by minimizing the average energy in the state 
$|\vartheta\rangle$ we again recover the Gross-Pitaevskii equation. More 
important for our purposes is that we can easily show by Fourier 
analysis that
\begin{equation}
|N\rangle =
  \int_0^{2\pi} \frac{d\vartheta}{2\pi}~ e^{-iN\vartheta} |\vartheta\rangle~,
\end{equation}
which shows that the number of particles $N$ and the phase $\vartheta$ are 
conjugate variables and obey the commutation relation $[N,\vartheta] = i$. The 
Heisenberg equation of motion for the average phase thus becomes
\begin{equation}
i\hbar \frac{\partial \langle\vartheta\rangle}{\partial t} 
  = \left\langle [\vartheta,H] \right\rangle 
  = -i \left\langle \frac{\partial H}{\partial N} \right\rangle \equiv -i \mu~,
\end{equation}
with $\mu$ the chemical potential of the gas. We therefore recover the important 
Josephson relation
\begin{equation}
\label{JR}
\hbar \frac{\partial \langle\vartheta\rangle}{\partial t} = -\mu~,
\end{equation}
which is also well known from the hydrodynamic formulation of the 
Gross-Pitaevskii equation \cite{sandro}. Indeed, taking the gradient of this 
equation and using the definition of the superfluid velocity, i.e., 
${\bf v}_s = \hbar \mbox{\boldmath $\nabla$} \langle\vartheta\rangle/m$, we 
obtain the desired result that
\begin{equation}
\label{vs}
\frac{\partial {\bf v}_s}{\partial t} 
        = - \frac{1}{m} \mbox{\boldmath $\nabla$} \mu~.
\end{equation}

Let us now return to Cooper-pair formation in degenerate Fermi gases. In analogy 
with Bose-Einstein condensation we can now use at zero temperature the 
variational wave function
\begin{equation}
\label{fixedN}
|N\rangle = \frac{ \left( \displaystyle \int d{\bf x} \int d{\bf x}'~ 
       \phi_{\alpha,\alpha'}({\bf x},{\bf x}')
       \psi_{\alpha'}^{\dagger}({\bf x}') \psi_{\alpha}^{\dagger}({\bf x})
                                        \right)^{N/2}}{(N/2)!}~|0\rangle~.
\end{equation}
A few remarks are in order. First, we have denoted the various hyperfine states 
of the atoms by $|\alpha\rangle$. In our previous discussion of Bose-Einstein 
condensation we should in principle also have indicated the hyperfine state of 
the atoms and used $\psi^{\dagger}_{\alpha}({\bf x})$ instead of 
$\psi^{\dagger}({\bf x})$ in the variational wave function. However, as long as 
all the atoms are in the same hyperfine state, the precise atomic state which is 
trapped is unimportant from a theoretical point of view and only influences the 
interatomic interaction, i.e., the particular value of the scattering length, 
that should be used in the Gross-Pitaevskii equation. For clarity we therefore 
suppressed the spin degrees of freedom in that case. 

Second, the fermionic creation operators anticommute. As a result the 
Cooper-pair wave function must obey 
$\phi_{\alpha,\alpha'}({\bf x},{\bf x}') = 
                                - \phi_{\alpha',\alpha}({\bf x}',{\bf x})$, 
reflecting the Pauli exclusion principle. There are essentially two ways to 
fulfill this antisymmetrization requirement. If all the atoms are in the same 
hyperfine state we have $\alpha'=\alpha$ and the orbital part of the Cooper-pair 
wave function is antisymmetric with respect to an exchange of the atoms. The 
relative angular momentum of the pairs must therefore be odd. In particular, we 
can have $p$-wave pairing. This situation occurs in doubly spin-polarized Fermi 
gases and also in liquid $^3$He \cite{tony1}. If we have an equal number of atoms 
in two hyperfine states, which is implicitly assumed in the above variational 
wave function, the spin part of the Cooper-pair wave function can already be 
antisymmetric and the relative angular momentum of the pairs must then be even. 
We now can have $s$-wave pairing as in ordinary superconductors. In principle, 
we can of course also have $d$-wave pairing as in the high-temperature 
superconductors, but this turns out to be extremely unlikely for dilute gases.   
                                
Third, in actual applications it is again more convenient to use a variational 
wave function with a well defined phase. In this case it is obtained by 
multiplying the Cooper-pair wave function in the right-hand side of 
Eq.~(\ref{fixedN}) with $e^{2i\vartheta}$ and summing over all even values of 
$N$. We then find
\begin{equation}
|\vartheta\rangle =
  \exp \left( \int d{\bf x} \int d{\bf x}'~ 
         \phi_{\alpha,\alpha'}({\bf x},{\bf x}') e^{2i\vartheta}
         \psi_{\alpha'}^{\dagger}({\bf x}') \psi_{\alpha}^{\dagger}({\bf x})
       \right) |0\rangle
\end{equation}
and exactly the same relation between the states $|N\rangle$ and 
$|\vartheta\rangle$ as for a condensate of single atoms. The Josephson relation 
given in Eq.~(\ref{JR}) is therefore also valid in this case. Moreover, in the 
state $|\vartheta\rangle$ we have a nonvanishing expectation value
\begin{equation}
\langle \vartheta |\psi_{\alpha}({\bf x}) \psi_{\alpha'}({\bf x}')  
  |\vartheta\rangle = \phi_{\alpha,\alpha'}({\bf x},{\bf x}') e^{2i\vartheta}~,
\end{equation}
which suggest that 
$\langle \psi_{\alpha}({\bf x}) \psi_{\alpha'}({\bf x}') \rangle$ is the 
appropriate order parameter for the Bardeen-Cooper-Schrieffer transition, just 
like $\langle \psi_{\alpha}({\bf x}) \rangle$ is the order parameter for 
Bose-Einstein condensation. Although this identification of the order parameter 
is correct, it turns out that it is more convenient in practice to work with the 
so-called BCS gap parameter
\begin{equation}
\Delta({\bf r};{\bf R}) 
  \equiv V_{\alpha,\alpha'}({\bf r}) 
               \langle \psi_{\alpha}({\bf R}+{\bf r}/2) 
                       \psi_{\alpha'}({\bf R}-{\bf r}/2) \rangle~,
\end{equation}
where $V_{\alpha,\alpha'}({\bf r}) = 
          \langle \alpha,\alpha'|V^c(r) + V^d({\bf r}) | \alpha,\alpha' \rangle$ 
is a shorthand notation for the elastic part of the interatomic interaction, and ${\bf r} = {\bf x} - {\bf x}'$ and ${\bf R} = ({\bf x} + {\bf x}')/2$ are the relative and center-of-mass coordinates of the Cooper pair, respectively. 
                  
Knowing the order parameter of the phase transition of interest, it is 
straightforward to obtain the corresponding mean-field theory. The main idea is 
first to write in the interaction part of the hamiltonian
\begin{eqnarray}
H &=& \sum_{\alpha} \int d{\bf x}~
      \psi_{\alpha}^{\dagger}({\bf x})
         \left( - \frac{\hbar^2 \mbox{\boldmath $\nabla$}^2}{2m} 
                + V^{\rm trap}({\bf x}) - \mu_{\alpha}
         \right) \psi_{\alpha}({\bf x})                        \\
  &+& \frac{1}{2} \sum_{\alpha,\alpha'} \int d{\bf x} \int d{\bf x}'~
         \psi_{\alpha'}^{\dagger}({\bf x}') \psi_{\alpha}^{\dagger}({\bf x})
            V_{\alpha,\alpha'}({\bf x}-{\bf x}')
               \psi_{\alpha}({\bf x}) \psi_{\alpha'}({\bf x}') \nonumber
\end{eqnarray}
the operators $\psi_{\alpha}({\bf x}) \psi_{\alpha'}({\bf x}')$ and 
$\psi_{\alpha}^{\dagger}({\bf x}) \psi_{\alpha'}({\bf x}')$ as a sum of a mean 
value and fluctuations, and then to neglect terms that are quadratic in the 
fluctuations. In this manner we arrive at the mean-field hamiltonian
\begin{eqnarray}
~~~~~H_{MF} &=& 
     \sum_{\alpha} \int d{\bf x}~
      \psi_{\alpha}^{\dagger}({\bf x})
         \left( - \frac{\hbar^2 \mbox{\boldmath $\nabla$}^2}{2m} 
                + V^{\rm trap}({\bf x}) - \mu'_{\alpha}({\bf x})
         \right) \psi_{\alpha}({\bf x})                       \\
  &+& \frac{1}{1+\delta_{\alpha,\alpha'}} \int d{\bf x} \int d{\bf x}'~
         \left( \Delta({\bf r};{\bf R}) 
                \psi^{\dagger}_{\alpha'}({\bf x}')  
                \psi^{\dagger}_{\alpha}({\bf x})  +
                \Delta^*({\bf r};{\bf R})
                \psi_{\alpha}({\bf x}) 
                \psi_{\alpha'}({\bf x}')  
         \right)~,                                              \nonumber
\end{eqnarray}
where the renormalized chemical potentials are in a good approximation (but see 
below) given by
\begin{equation}
\label{chem}
\mu'_{\alpha}({\bf x}) = 
  \mu_{\alpha} - \sum_{\alpha' \neq \alpha} \int d{\bf x}'~
     V_{\alpha,\alpha'}({\bf x}-{\bf x}') n_{\alpha'}({\bf x'})
\end{equation}
and the density profile of atoms in spin state $|\alpha\rangle$ obeys
$n_{\alpha}({\bf x}) = \langle \psi_{\alpha}^{\dagger}({\bf x})
                               \psi_{\alpha}({\bf x}) \rangle$. It is important 
to note that in trapped alkali gases it is indeed appropriate to have a chemical 
potential for each spin state, because the time scale for relaxation towards 
equilibrium in spin space is always much larger than the equilibration time for 
the spatial degrees of freedom. This is for example quite dramatically 
demonstrated by the two overlapping condensate experiments by Myatt \etal 
\cite{JILA2}.
                               
To complete the mean-field theory we should now calculate the mean values of the 
operators $\psi_{\alpha}({\bf x}) \psi_{\alpha'}({\bf x}')$ and 
$\psi_{\alpha}^{\dagger}({\bf x}) \psi_{\alpha'}({\bf x}')$ in a thermal 
ensemble with the hamiltonian $H_{MF}$. This clearly makes the theory 
selfconsistent. To perform the calculation we write the annihilation operators 
at the positions ${\bf x}$ and ${\bf x}'$ as
\begin{equation}
\psi_{\alpha}({\bf R} \pm {\bf r}/2) 
  = \int \frac{d{\bf k}}{(2\pi)^3}~ \psi_{\alpha}({\bf k}) 
                                 e^{i{\bf k} \cdot ({\bf R} \pm {\bf r}/2)}~,
\end{equation}
respectively. Substituting this in the mean-field hamiltonian and neglecting 
gradients in the center-of-mass coordinate ${\bf R}$, we find that
\begin{eqnarray}
~~~~~H_{MF} &=& 
     \sum_{\alpha} \int \frac{d{\bf k}}{(2\pi)^3}~
      \psi_{\alpha}^{\dagger}({\bf k})
         \left( - \frac{\hbar^2 {\bf k}^2}{2m} 
                + V^{\rm trap}({\bf R}) - \mu'_{\alpha}({\bf R})
         \right) \psi_{\alpha}({\bf k})                       \\
  &+& \frac{1}{1+\delta_{\alpha,\alpha'}} \int \frac{d{\bf k}}{(2\pi)^3}~
         \left( \Delta({\bf k};{\bf R}) 
                \psi^{\dagger}_{\alpha'}(-{\bf k})  
                \psi^{\dagger}_{\alpha}({\bf k})  +
                \Delta^*({\bf k};{\bf R})
                \psi_{\alpha}({\bf k}) 
                \psi_{\alpha'}(-{\bf k})  
         \right)~.                                              \nonumber
\end{eqnarray}
Diagonalizing the above hamiltonian by means of a Bogoliubov transformation, the 
spin-density profiles can then be calculated from
\begin{equation}
\label{spin}
n_{\alpha}({\bf R}) = \int \frac{d{\bf k}}{(2\pi)^3}~
    \langle \psi_{\alpha}^{\dagger}({\bf k}) \psi_{\alpha}({\bf k}) \rangle 
\end{equation}
and, most importantly, the BCS gap parameter from
\begin{equation}
\label{gap}
\Delta({\bf k};{\bf R}) = \int \frac{d{\bf k}'}{(2\pi)^3}~
  V_{\alpha,\alpha'}({\bf k}-{\bf k}') 
     \langle \psi_{\alpha}({\bf k}') \psi_{\alpha'}(-{\bf k}') \rangle~.
\end{equation}             
Notice that the averages in the right-hand side of Eqs.~(\ref{spin}) and 
(\ref{gap}) depend on the position ${\bf R}$, since the mean-field hamiltonian 
$H_{MF}$ depends parametrically on this position. To arrive at such a simplyfied 
(Thomas-Fermi) description of the inhomogeneity of the gas we have to be able to 
neglect gradients of the densities $n_{\alpha}({\bf R})$ and the BCS gap 
parameter $\Delta({\bf k};{\bf R})$. This is indeed true for present experiments 
with fermionic alkali gases, because the number of trapped atoms is so large 
that both the correlation length and the size of the Cooper pairs are small 
compared the typical length scale on which the trapping potential varies 
\cite{marianne1,baranov}. 

We refer for the details of the diagonalization to our previous work 
\cite{marianne1}. For our present purposes it is however important to mention 
that in the optimal case of equal spin densities, the final result of 
Eq.~(\ref{gap}) is the famous BCS gap equation \cite{BCS,tony1}
\begin{equation}
\Delta({\bf k};{\bf R}) = - \int \frac{d{\bf k}'}{(2\pi)^3}~
  V_{\alpha,\alpha'}({\bf k}-{\bf k}')
    \frac{\Delta({\bf k}';{\bf R})}{2\hbar\omega({\bf k}';{\bf R})} 
       {\rm tanh}\left( \frac{\hbar\omega({\bf k}';{\bf R})}{2k_BT} \right)~,
\end{equation}
where the so-called Bogliubov dispersion $\hbar\omega({\bf k};{\bf R})$ obeys
\begin{equation} 
\hbar\omega({\bf k};{\bf R}) =
   \sqrt{ \left( \frac{\hbar^2{\bf k}^2}{2m} - \epsilon_F({\bf R}) \right)^2
                          + |\Delta({\bf k};{\bf R})|^2 }~.
\end{equation}
In principle, one can solve the BCS equation for any interatomic potential 
$V_{\alpha,\alpha'}({\bf r})$. However, for alkali gases the complete central 
interaction is usually not very well known and we have only information on the 
two-body scattering length $a$. We therefore would like to reformulate the gap 
equation in such a way that only this scattering length enters. This is achieved 
by noting that the gap equation is very similar to the Lippmann-Schwinger 
equation for the two-body T(ransition) matrix. Indeed the latter reads 
\cite{henk3}
\begin{equation}
~~~~~~T_{\alpha,\alpha'}^{2B}({\bf k},{\bf k}'';z) = 
  V_{\alpha,\alpha'}({\bf k}-{\bf k}'') +
  \int \frac{d{\bf k}'}{(2\pi)^3}~ V_{\alpha,\alpha'}({\bf k}-{\bf k}')
    \frac{1}{z-\hbar^2{\bf k}'^2/m} 
                            T_{\alpha,\alpha'}^{2B}({\bf k}',{\bf k}'';z)~.
\end{equation} 
Using the Lippman-Schwinger equation with $z=2\epsilon_F({\bf R}) + i0$, we can 
after some algebraic manipulation show that the gap equation is equivalent to
\begin{eqnarray}
\Delta({\bf k};{\bf R}) = - \int \frac{d{\bf k}'}{(2\pi)^3}~
 T_{\alpha,\alpha'}^{2B}({\bf k},{\bf k}';2\epsilon_F({\bf R}))
  \left( \frac{1}{2\hbar\omega({\bf k}';{\bf R})} 
          {\rm tanh}\left( \frac{\hbar\omega({\bf k}';{\bf R})}{2k_BT} \right)
  \right.                                                            \\
  \left.    - \frac{1}{\hbar^2{\bf k}'^2/m - 2\epsilon_F({\bf R}) - i0}
  \right) \Delta({\bf k}';{\bf R})~.                                 \nonumber
\end{eqnarray} 
This result serves our purposes since the two-body T matrix is directly related 
to the two-body scattering length as we will see next.

\section{Doubly spin-polarized Fermi gases}
In doubly spin-polarized Fermi gases all the atoms are in the same hyperfine 
state. Although it is straightforward to generalize the following to an 
arbitrary state $|\alpha\rangle$, we will in first instance restrict ourselves 
to the fully stretched state $|\zeta\rangle$ in which both the electron as well 
as the nuclear spin have a maximal projection on the magnetic field axis. The  
relevant interaction matrix element is then equal to 
\begin{equation}
V_{\zeta,\zeta}({\bf r}) = V_T(r) +
    \frac{\mu_0\mu_e^2}{4\pi r^3} \left( 1 - 3 \cos^2(\theta_{\bf r}) \right)~,
\end{equation} 
with $\theta_{\bf r}$ the angle between the interatomic separation ${\bf r}$ and 
the magnetic field ${\bf B}$. For this potential we have to calculate the 
two-body T matrix. Treating the weak electron-electron magnetic dipole 
interaction in Born approximation \cite{henk2} and making use of the fact that 
for the central interaction only $p$-waves contribute at low momenta, we find
\begin{equation}
T_{\zeta,\zeta}^{2B}({\bf k},{\bf k}';z) =
  \frac{12\pi a^3 \hbar^2}{m} {\bf k} \cdot {\bf k'} 
  + \frac{\mu_0\mu_e^2}{8\pi} 
               \left( \cos^2(\theta_{{\bf k} - {\bf k}'}) 
                      - \cos^2(\theta_{{\bf k} + {\bf k}'})\right)~.  
\end{equation}
Here $a$ is the triplet $p$-wave scattering length and 
$\theta_{{\bf k} \pm {\bf k}'}$ denotes the angle between the momentum transfer 
${\bf k} \pm {\bf k}'$ and the magnetic field axis.

We now have two cases to consider. Generically we expect the $p$-wave scattering 
length to be of the order of the range of the triplet potential and therefore 
$k_F({\bf R})|a|$ to be much smaller than one. In that case the contribution of the triplet potential to the two-body T matrix is negligible and the effective 
interaction between the atoms is dominated by the long-range dipole-dipole 
interaction. Due to the complicated angular dependence of this interaction it is 
not possible to solve the BCS gap equation analytically. However, we can 
nevertheless make progress by noting that the magnetic dipole-dipole interaction is most attractive when ${\bf r}$ is directed along the magnetic field. We thus expect that if Cooper pairs are formed their wavefunction 
$\phi_{\zeta,\zeta}({\bf r};{\bf R})$ is most likely proportional to 
$Y_{1,0}({\bf \hat{r}}) = \sqrt{3/4\pi} \cos(\theta_{\bf r})$, which implies 
that $\Delta({\bf k};{\bf R}) = Y_{1,0}({\bf \hat{k}}) \Delta({\bf R})$. Since 
this gap is anisotropic in the relative wave vector ${\bf k}$, the gas is below 
the critical temperature an anisotropic superfluid, just like $^3$He in the 
so-called A phases \cite{tony1}. 

To obtain an estimate for the critical temperature we explicitly consider only 
the $p$-wave part of the dipole-dipole interaction, which results in the 
approximation 
\begin{equation}
T_{\zeta,\zeta}^{2B}({\bf k},{\bf k}';z) \simeq
 - \frac{4}{15} \pi \mu_0\mu_e^2 
       \sum_{m} (-1)^m (1+\delta_{m,0}) 
           Y_{1,m}({\bf \hat{k}}) Y^*_{1,m}({\bf \hat{k}}')~.
\end{equation}
This explicitly confirms that the dipole-dipole interaction is only attractive 
in the channel $m=0$. Substituting the above two-body T matrix into the BCS gap 
equation and linearizing with respect to $\Delta({\bf 0})$ to obtain an equation 
for the critical temperature, we find 
\begin{equation}
\frac{2}{15 \mu_0\mu_e^2} = 
 \int \frac{d{\bf k}}{(2\pi)^3}~
      \frac{\cal P}{\epsilon_F({\bf 0}) - \hbar^2{\bf k}^2/2m}
         N({\bf k};{\bf 0})~.
\end{equation} 
Here we used the notations ${\cal P}$ for the Cauchy principle value part of the 
integral and $N({\bf k};{\bf R})$ for the Fermi distribution function 
$N(\epsilon) = (e^{\beta \epsilon} + 1)^{-1}$ evaluated at 
$\hbar^2{\bf k}^2/2m - \epsilon_F({\bf R})$. Note that we have also used that 
the BCS gap equation will have a nontrivial solution in the center of the 
trap first, since the density of the gas is highest there. Introducing a `scattering length' $a^d$ for the electron-electron magnetic dipole interaction by means of $4\pi a^d \hbar^2/m \equiv -15 \mu_0\mu_e^2/2$, i.e., 
$a^d = - 15 \mu_0\mu_e^2 m/8\pi \hbar^2 \simeq - 0.2 (m/m_H)$ a$_0$ with $m_H$ the mass of a hydrogen atom and a$_0$ its Bohr radius, the equation for the critical temperature becomes identical to the linearized BCS gap equation that has been studied previously in the context of $s$-wave superconductors by S\'a de Melo \etal \cite{mohit}. Using their result, we have for the critical temperature
\begin{equation}
T_c = \frac{8\epsilon_F({\bf 0}) e^{\gamma-2}}{k_B \pi} 
         \exp \left\{ - \frac{\pi}{2k_F({\bf 0})|a^d|} \right\}~,
\end{equation}
with $\gamma \simeq 0.5772$ Euler's constant. Unfortunately, the BCS transition 
to an anisotropic superfluid thus occurs at extremely low temperatures in this 
case and appears to be out of reach experimentally. For example for $^6$Li at a 
density of $n \simeq 1 \times 10^{12}$ cm$^{-3}$, we have $\epsilon_F/k_B \simeq 
600$ nK and $k_F|a^d| \simeq 2 \times 10^{-4}$.

The second case to consider appears to be more promising. As mentioned before, 
the triplet $p$-wave scattering length is in general too small to be able to 
dominate over the dipole-diple interaction. For $^6$Li it is only $-35$ a$_0$, 
for instance. However, we can imagine that it is possible, in the same way as in 
the recent experiment by Inouye \etal \cite{MIT3}, to (optically) trap a 
hyperfine state that has a $p$-wave Feshbach resonance \cite{eite} and tune the 
external bias magnetic field such that the $p$-wave scattering length becomes 
large and negative. The two-body T matrix is then well approximated by
\begin{equation}
T_{\alpha,\alpha}^{2B}({\bf k},{\bf k}';z) \simeq
   \frac{12\pi a^3 \hbar^2}{m} {\bf k} \cdot {\bf k'} =
   \frac{(4\pi)^2 a^3 \hbar^2}{m} k k'
       \sum_{m} Y_{1,m}({\bf \hat{k}}) Y^*_{1,m}({\bf \hat{k}}')~.
\end{equation} 
Substituting this into the BCS gap equation, it is not difficult to show that it 
is solved by the {\it ansatz} 
\begin{equation}
\Delta({\bf k};{\bf R}) 
    = k \Delta({\bf R}) \sum_m d_m^* Y_{1,m}({\bf \hat{k}})
    = \sqrt{\frac{3}{4\pi}} \Delta({\bf R}) {\bf d} \cdot {\bf k}~,
\end{equation}
where $d_m$ are the spherical components of a vector ${\bf d}$ that is normalized as $\sum_m |d_m|^2 = {\bf d} \cdot {\bf d} = 1$. Furthermore, linearizing the BCS gap equation with respect to $\Delta({\bf 0})$, we find that the critical temperature is now determined by
\begin{equation}
- \frac{m}{4\pi a^3 \hbar^2} = 
  \int \frac{d{\bf k}}{(2\pi)^3}~
       {\bf k}^2 \frac{\cal P}{\epsilon_F({\bf 0}) - \hbar^2{\bf k}^2/2m}
         N({\bf k};{\bf 0})~.
\end{equation} 
The solution can be obtained by the same methods as before and reads
\begin{equation}
T_c = \frac{8\epsilon_F({\bf 0}) e^{\gamma - 8/3}}{k_B\pi}  
            \exp \left\{ - \frac{\pi}{2(k_F({\bf 0})|a|)^3} \right\}~.
\end{equation}

It is important to realize that up to this point the direction of the vector 
${\bf d}$ is arbitrary. This reflects the rotational symmetry of the problem. 
However, as we have seen, the magnetic dipole-dipole interaction breaks this 
symmetry and will cause {\bf d} to lie either parallel or perpendicular to the 
magnetic field, depending on precisely which hyperfine state $|\alpha\rangle$ is 
trapped. Because we are dealing with an attractive interaction we also
have to make sure that the gas is mechanically stable. This leads to the 
restriction that $n({\bf 0})|a|^3 \le 5/96\pi$, or equivalently 
$(k_F({\bf 0})|a|)^3 \le 5/576\pi^3$. Unfortunately, the latter again severely limits the feasibility of experimentally observing the in principle interesting possibility of a transition to an anisotropic superfluid.  

\section{Spin-polarized Fermi gases}
We now turn our attention to the case of fermionic gases that are a mixture of two hyperfine states. In such a gas we can have $s$-wave collisions between atoms in different hyperfine states, which has as an advantage that interacting effects are expected to be much more important. Moreover, it has an additional advantage that it is now in principle possible to use evaporative cooling to cool the gas to low temperatures. A fermionic gas consisting of three hyperfine states has recently been considered as well \cite{tony2}, but since it does not lead to any qualitative different physics we restrict ourselves here to mixtures of only two spin states. We denote the two hypefine states involved from now on by $|\uparrow\rangle$ and $|\downarrow\rangle$. Experimentally, there are two ways to realize such a system. We can either magnetically trap two low-field seeking states, or use an optical trap. In the latter case it is presumably necessary to load the trap by precooling the gas in a magnetic trap using sympathetic cooling.   
 
\subsection{Positive scattering length}
If the $s$-wave scattering length $a$ between two atoms in different spin states is positive, a first many-body effect that we have to consider is the phase separation of the gas into two phases with opposite `magnetization'. Roughly speaking, this implies that instead of having overlapping spin densities $n_{\uparrow}({\bf R})$ and $n_{\downarrow}({\bf R})$, the gas prefers to separate into two phases in which (almost) all the atoms are either in the state $|\uparrow\rangle$ or in the state $|\downarrow\rangle$. The driving force behind this instability is that although the phase separation increases the kinetic energy of the gas, this increase is more than compensated by the decrease in interaction energy. More precisely, the gas is stable if the free-energy surface $F[n_{\uparrow},n_{\downarrow}]$ has a positive curvature in all directions. Using that 
\begin{equation}
\label{T2B}
T_{\uparrow,\downarrow}^{2B}({\bf k},{\bf k}';z) = \frac{4\pi a \hbar^2}{m}~, 
\end{equation} 
the free energy in the degenerate regime is given by
\begin{equation}
~~~~~F[n_{\uparrow},n_{\downarrow}] = \int d{\bf R}~ 
 \left\{
  (6\pi^2)^{2/3} \frac{3\hbar^2}{10m} 
     \left( n_{\uparrow}^{5/3}({\bf R}) + n_{\downarrow}^{5/3}({\bf R}) \right) 
        + \frac{4\pi a \hbar^2}{m} n_{\uparrow}({\bf R}) n_{\downarrow}({\bf R})
 \right\}
\end{equation}
and stability of the gas requires that the spin densities obey
$n_{\uparrow}({\bf 0}) n_{\downarrow}({\bf 0}) a^6 \le (\pi/48)^2$. In the particular case of equal spin densities this reduces to the condition that $k_F({\bf 0})a \le \pi/2$. Note that above we have used the T-matrix approximation to evaluate the average interaction energy. Because the chemical potential $\mu_{\alpha}$ is equal to the derivative 
$\delta F/\delta n_{\alpha}({\bf R})$, we must for consistency use the same approximation to determine the renormalized chemical potentials $\mu'_{\alpha}({\bf R})$. Therefore, we use in the following always that
\begin{equation}
\mu'_{\alpha}({\bf R}) = 
  \mu_{\alpha} - \sum_{\alpha' \neq \alpha} \frac{4\pi a \hbar^2}{m}
                                                       n_{\alpha'}({\bf R})
\end{equation}
instead of the less accurate (only Born approximation) relation given in Eq.~(\ref{chem}).

Even though the interaction between the atoms is effectively repulsive, there can nevertheless occur a BCS pairing transition to a superfluid state due to the so-called Luttinger-Kohn instability \cite{kohn}. Physically, the instability is a result of the fact that two atoms in the same hyperfine state can exchange a fluctuation (phonon) in the density of the other hyperfine state, leading to an effectively attractive interaction between the atoms involved. The $p$-wave transition associated with the Luttinger-Kohn effect has been studied by Baranov \etal \cite{kagan1}. These authors obtain for the critical temperature the estimate
\begin{equation}
T_c \simeq \frac{\epsilon_F({\bf 0})}{k_B}  
            \exp \left\{ - 13 \left(\frac{\pi}{2k_F({\bf 0})a} \right)^2
                 \right\}~,
\end{equation}
where notably $a$ is the $s$-wave scattering length.
It should, however, be kept in mind that to be able to observe the transition we must require that the gas is mechanically stable and does not phase separate. As we have already seen, this implies for equal spin densities that 
$k_F({\bf 0})a \le \pi/2$. Moreover, for generic values of the scattering length we will even have that $k_F({\bf 0})a \ll 1$ at the low densities of interest. It therefore seem once again practically impossible to observe the above transition in a real atomic gas. 

\subsection{Negative scattering length}
Our last chance of achieving a superfluid phase in a fermionic gas thus appears to be a spin-polarized gas with a large and negative $s$-wave scattering length. Besides the possibility of using a Feshbach resonance to tailor the scattering length, we can also directly make use of the anomalously large $^6$Li triplet scattering length of $-2160$ a$_0$ by trapping a spin-polarized $^6$Li gas in a bias magnetic field of at least $0.05$ T \cite{marianne2}. Substituting Eq.~(\ref{T2B}) into the BCS gap equation we see that the solution 
$\Delta({\bf k};{\bf R})$ is now independent of the wave vector ${\bf k}$ and therefore equal to $\Delta({\bf R})$. Furthermore, a linearization in $\Delta({\bf R})$ gives 
\begin{equation}
- \frac{m}{4\pi a \hbar^2} = 
 \int \frac{d{\bf k}}{(2\pi)^3}~
      \frac{\cal P}{\epsilon_F({\bf 0}) - \hbar^2{\bf k}^2/2m}
         N({\bf k};{\bf 0})~,
\end{equation} 
which results in the critical temperature 
\begin{equation}
T_c = \frac{8\epsilon_F({\bf 0}) e^{\gamma-2}}{k_B \pi} 
         \exp \left\{ - \frac{\pi}{2k_F({\bf 0})|a|} \right\}~.
\end{equation}
Since the mechanical stability of the gas requires also for a negative scattering length only that $k_F({\bf 0})|a| \le \pi/2$, this expression shows that the prospects of observing a BCS transition in this case are indeed most favorable. 

\begin{figure}[p]
\vspace*{2.5in}
\hspace*{-0.2in}
\psfig{figure=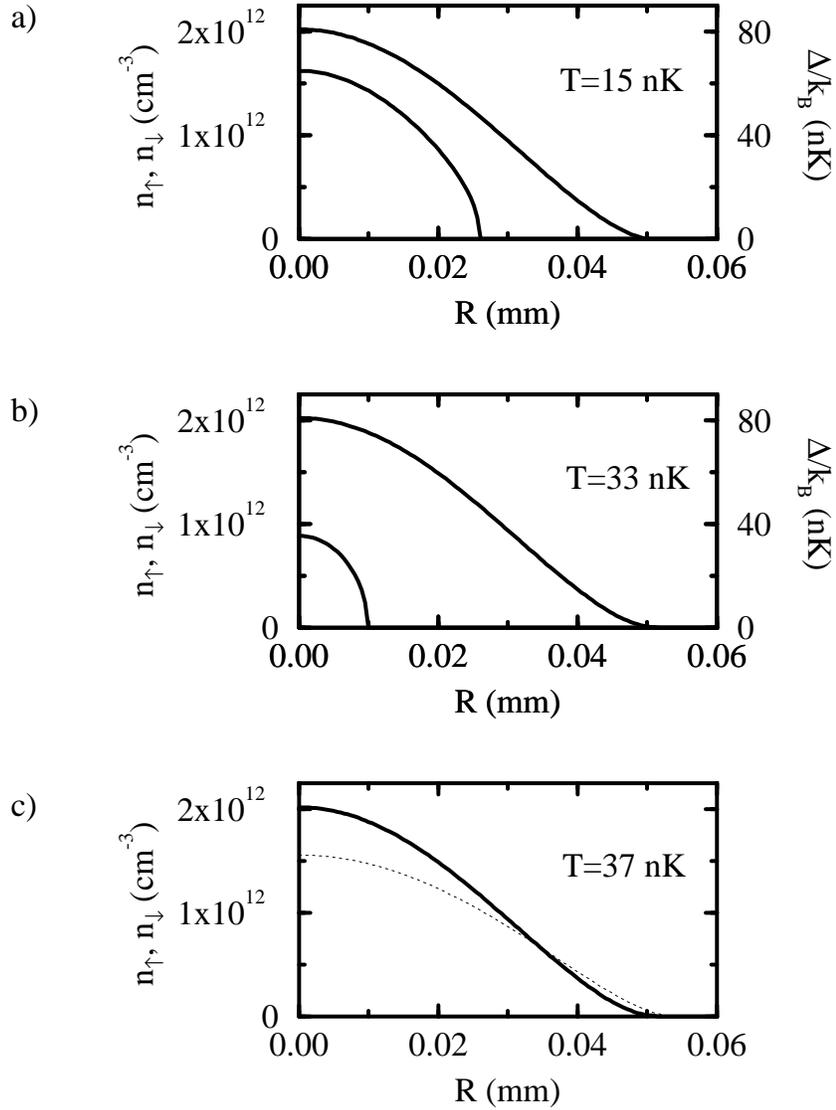}
\caption{Density distribution $n_{\uparrow}({\bf R}) = n_{\downarrow}({\bf R})$
         and energy gap $\Delta({\bf R})$ for a $^6$Li atomic gas consisting
         of $2.865 \times 10^5$ atoms in each spin state a) at $T=15$ nK,
         b) at $T=33$ nK, slightly below $T_c$, and c) at $T=T_c=37$ nK.
         The left scale of each plot refers to the density and the right scale
         to the energy gap. The dotted line in c) shows the density 
         distribution for an ideal Fermi gas with the same number of 
         particles and at the same temperature.
         \label{profiles}}
\end{figure}

\begin{figure}[hb]
\vspace*{1.5in}
\psfig{figure=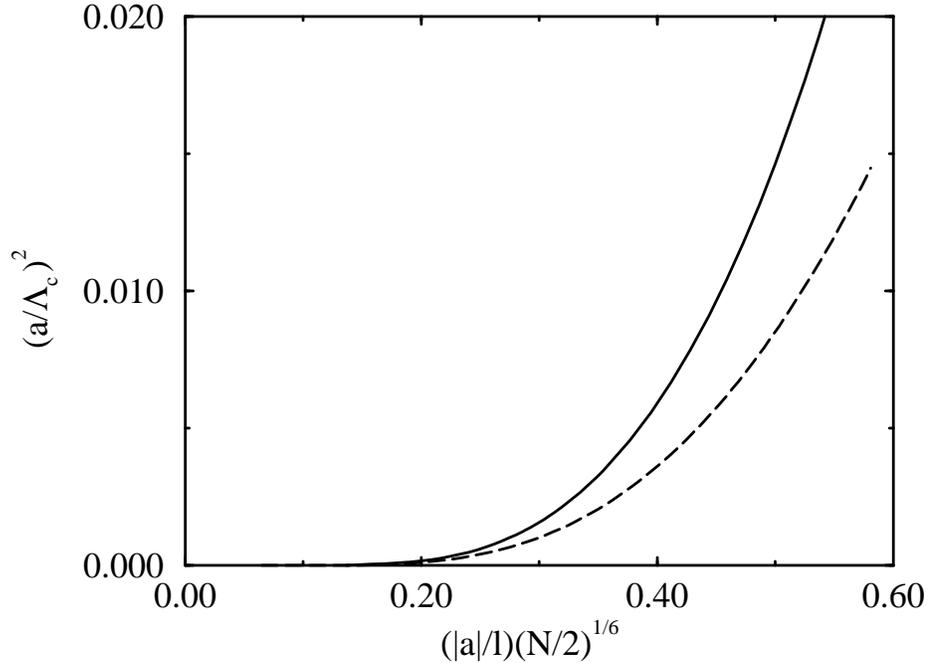}
\vspace*{-0.2in}
\caption{Critical temperature as a function of the number of particles
         (solid line) when there are $N/2$ particles present in both spin
         states. The dashed line represents the critical temperature for a gas
         whose density distribution is not altered by mean-field interactions.
         \label{Tc}}
\end{figure}

In view of this encouraging situation we have studied in more detail the behavior of the spin density profile $n_{\alpha}({\bf R})$ and the BCS gap parameter $\Delta({\bf R})$ for a spin-polarized $^6$Li gas in the same magnetic trap that has been used for the Bose-Einstein condensation experiments with $^7$Li \cite{Rice1,Rice2}. The results for equal spin densities are shown in Fig.~\ref{profiles} and lead to three important conclusions.    
First, we see explicitly that the density profile of a degenerate Fermi gas is indeed completely `frozen'. Clearly, even the BCS transition to a superfluid has essentially no effect on the density profile. From an experimental point of view this is somewhat unfortunate, because it implies that the appearance of a condensate of Cooper pairs cannot be observed by the same methods that have been so successful in the Bose-Einstein condensation experiments. We come back to this problem shortly. Second, for a total density of $n \simeq 4 \times 10^{12}$ cm$^{-3}$ the critical temperature is about $37$ nK. In view of the achievements with bosonic alkali gases, this appears to be a density-temperature combination which is certainly within reach experimentally. Finally, we have also compared the density profile of $^6$Li with that of a noninteracting Fermi gas with the same number of particles. The strong attractive interaction between the $^6$Li atoms evidently results in a substantial increase of the density in the center of the trap. Experimentally, this is a favorable effect because for a fixed number of atoms in the trap it enhances the critical temperature as is shown quantitatively in Fig.~\ref{Tc}. 
For future convenience we mention that in a good approximation the critical temperature obeys 
\begin{equation}
\left( \frac{a}{\Lambda_c} \right)^2 \simeq
  0.037 \exp \left\{ 
               - 1.36 \frac{\ell}{|a| N^{1/6}} + 2.66 \frac{|a| N^{1/6}}{\ell}
             \right\}~,
\end{equation}
with $\Lambda = (2\pi\hbar^2/mk_BT)^{1/2}$ the thermal de Broglie wavelength  and $\ell = (\hbar/m\omega)^{1/2}$ the size of the harmonic oscillator ground state. It must be kept in mind that the latter formula can only be used for values of $N^{1/6}|a|/\ell$ that are less then about $0.74$, since for larger values the gas is mechanically unstable and undergoes a spinodal decomposition first.
     
\section{Discussion and conclusions}
We have argued that for condensed matter physics in trapped atomic Fermi gases, a spin-polarized gas with a large and negative $s$-wave scattering length between the atoms in the two different hyperfine states appears to be most promising. In view of the large uncertainties in the interatomic interaction potential of $^{40}$K, the most likely candidate for the achievement of a gaseous BCS superfluid is at present $^6$Li. However, before successful experiments with atomic $^6$Li can be performed, some experimental problems need to be resolved. One problem is that in spin-polarized atomic $^6$Li not only the $s$-wave scattering length but also the exchange and dipolar decay rates are anomalously large. As a result the gas has usually a very short lifetime. To enhance the lifetime to about $1$ s at the densities of interest, we can either apply a bias magnetic field of about $5$ T or use optical methods to trap two high-field seeking states. The latter solution seems to be most practical and is actively being persued at the moment. 

Assuming that we are able to achieve the necessary conditions for the BCS transition, the next problem that arises is the detection of the Cooper pair condensate. As we have seen, the density profile shows essentially no sign of the phase transition. Time-of-flight measurements, that were the `smoking gun' for the Bose-Einstein condensation experiments, are therefore not appropriate here. Another possible signature that comes to mind are the frequencies of the collective modes. Because of the large $s$-wave scattering length required for relatively high critical temperatures, the collective modes are always in the hydrodynamical regime. They are therefore described by the (local) conservation laws and the Josephson relation. In the case of equal spin densities we thus obtain the following set of equations. The continuity equation for a superfluid is
\begin{equation}
\frac{\partial n}{\partial t} = - \mbox{\boldmath $\nabla$} \cdot {\bf j}~,
\end{equation}
with the total density $n = n_n + n_s$ and the total current density 
${\bf j} = n_n {\bf v}_n + n_s {\bf v}_s$ consisting of a normal and superfluid contribution. In addition, Newton's law gives
\begin{equation}
\frac{\partial {\bf j}}{\partial t} = - \frac{1}{m}
  \left( \mbox{\boldmath $\nabla$} p 
                    + n \mbox{\boldmath $\nabla$} V^{\rm trap} \right)~,
\end{equation}
where $p$ denotes the pressure in the gas. Finally, we have also the Josephson relation, which leads to (cf. Eq.~(\ref{vs}))
\begin{equation}
\frac{\partial {\bf v}_s}{\partial t} 
        = - \frac{1}{m} \mbox{\boldmath $\nabla$} \mu~.
\end{equation}

In principle, we must also take into account the conservation of energy. However, for a degenerate Fermi gas the specific heats at constant pressure and volume are almost equal and the continuity equation for the total energy density essentially decouples from the previous ones. This has important consequences, because for density fluctuations the Josephson relation just copies Newton's law and we must conclude that the first sound modes are not affected by the BCS transition. Of course, a measurement of second sound modes would be a clear signature of the transition, but this is presumably quite hard experimentally because second sound is primarily a temperature wave due to the fact that the energy fluctuations are almost decoupled. In analogy with sound attenuation in superconductors, it has been suggested by Fetter \cite{fetter} that the damping of the first sound modes might be strongly influenced by the appearance of a Cooper pair condensate, but more work is needed to make sure whether this interesting suggestion would work.

A property of the gas that is certainly influenced by the BCS transition is the decay of the gas. Qualitatively this can be easily understood from the correlator method devised by Kagan \etal \cite{kagan2}. In this approach we can relate the decay rate constant $G({\bf R};T)$ for two-body decay to the rate constant in the normal phase $G({\bf R};T_c)$ by
\begin{eqnarray}
G({\bf R};T) &=& G({\bf R};T_c) 
 \frac{1}{n_{\uparrow}({\bf R}) n_{\downarrow}({\bf R})}
 \left\langle 
         \psi^{\dagger}_{\uparrow}({\bf R}) \psi^{\dagger}_{\downarrow}({\bf R})
         \psi_{\downarrow}({\bf R}) \psi_{\uparrow}({\bf R}) 
 \right\rangle \\
 &=& G({\bf R};T_c) \left( 1 + 
                       \frac{|\phi_{\uparrow,\downarrow}({\bf R},{\bf R})|^2}
                            {n_{\uparrow}({\bf R}) n_{\downarrow}({\bf R})}
                    \right)~.  \nonumber
\end{eqnarray}
Hence, the Cooper pair condensate enhances the decay of the gas. A quantitative estimate of the effect is somewhat complicated by the fact that we cannot use a pseudopotential to calculate the Cooper pair wavefunction $\phi_{\uparrow,\downarrow}({\bf R},{\bf R})$ from our knowledge of the gap parameter $\Delta({\bf R})$. In any case, the increase in the two-body decay rate of the gas can only be used as a detection method if the gas is trapped in a magnetic trap, since in an optical trap the two-body decay will essentially be eliminated and the lifetime of the gas is determined by the rate at which  photons scatter off the atoms in the gas. Because this is a one-atom problem, it is also not affected by the BCS transition.

A final detection method that we would like to mention is the scattering of a beam of $^6$Li atoms from the gas cloud. Since such an experiment is quite similar to a tunneling experiment, it is directly sensitive to the existence of  the gap parameter $\Delta({\bf R})$. Therefore, a measurement of the angular distribution of the scattered atoms appears to be a promising way to get detailed information about the condensate of Cooper pairs. Of course, to be most sensitive we need a very cold beam. However, this is clearly not an impossible requirement, because the first results of such experiments with a condensate of $^{87}$Rb atoms have recently been reported \cite{eric}. From a theoretical point of view, we are at present performing a study of the above scattering process to see how large the effects of the BCS transition are. It is interesting to note that it is also possible to detect vortices in this manner, because the atomic beam (in contrast to a laser beam) does not only see the core of the vortex but its complete velocity profile. This is in fact also true for a Bose condensate. An experiment of this sort may therefore also be of interest in the context of ongoing research on the properties of trapped atomic Bose gases.  

\acknowledgments
Most of the work presented here has been performed in close collaboration with 
Ian McAlexander, Cass Sackett, and Randy Hulet. We are very grateful for their 
continued encouragement and crucial contributions. We also thank Yvan Castin, 
Eric Cornell, Jean Dalibard, Allan Griffin, Massimo Inguscio, Tony Leggett, 
Andrei Ruckenstein, Guglielmo Tino and Peter Zoller for helpful discussions. We 
acknowledge support from the Stich\-ting Fundamenteel Onderzoek der Materie 
(FOM), which is financially supported by the Nederlandse Organisatie voor 
Wetenschappelijk Onderzoek (NWO).

\end{document}